\documentclass{article}

\usepackage[a4paper, total={6.05in, 9in}]{geometry}

\usepackage{graphicx} 
\usepackage[utf8]{inputenc}

\usepackage{amsmath}
\usepackage{amssymb}
\usepackage{amsfonts}
\usepackage{physics}
\usepackage{color}
\usepackage{xcolor}
\usepackage[]{hyperref}
\usepackage{etoolbox}
\usepackage{mathrsfs}
\usepackage{todonotes}
\usepackage{comment}
\usepackage{authblk}

\usepackage{csquotes}

\usepackage[
  backend=biber,
  style=phys,
  sorting=none,
  giveninits=true,
  articletitle=true,
  biblabel=brackets,
  pageranges=false,
  chaptertitle=false,
  doi=false,
  url=false,
  isbn=false,
  hyperref=true,
  natbib=true
]{biblatex}

\patchcmd{\section}
  {\centering}
  {\raggedright}
  {}
  {}
  
\patchcmd{\subsection}
  {\centering}
  {\raggedright}
  {}
  {}

\newcommand{\x}{\mathsf{x}}
\newcommand{\ee}{\mathrm{e}}
\newcommand{\W}{\mathcal{W}}
\newcommand{\F}{\mathcal{F}}
\newcommand{\intinf}{\int_{-\infty}^\infty}
\newcommand{\intoinf}{\int_{0}^\infty}
\newcommand{\eff}{\mathrm{eff\!}}

\DeclareMathOperator{\sinch}{\mathrm{sinch}}
\DeclareMathOperator{\sinc}{\mathrm{sinc}}

\DeclareMathOperator{\diag}{\mathrm{diag}}

\numberwithin{equation}{section}

\title{
Spin to win: enhancing the quantum Otto cycle with the circular Unruh effect
}

\author[1]{Leo J. A. Parry\thanks{leo.parry@nottingham.ac.uk}}
\author[2]{Dimitris Moustos\thanks{dimitris.moustos@newcastle.ac.uk}}
\author[3]{Benito A. Ju\'arez-Aubry\thanks{benito.juarezaubry@york.ac.uk}}
\author[2]{Obinna Abah\thanks{obinna.abah@newcastle.ac.uk}}
\author[1]{Jorma Louko\thanks{jorma.louko@nottingham.ac.uk}}

\affil[1]{School of Mathematical Sciences,
University of Nottingham, \break Nottingham NG7 2RD, United Kingdom}
\affil[2]{School of Mathematics, Statistics, and Physics, Newcastle University, \break Newcastle upon Tyne NE1 7RU, United Kingdom}
\affil[3]{Department of Mathematics, Ian Wand Building, Deramore Lane, \break University of York,
York YO10 5GH, United Kingdom}

\date{November 2026}

\begin{document}

\maketitle

\begin{abstract}
We study a quantum Otto cycle whose working medium is a qubit Unruh-DeWitt detector undergoing uniform circular motion in $(3+1)$-dimensional Minkowski spacetime, interacting with a massless scalar field in a thermal state during the isochoric strokes. We compute the detector's transition rate and determine its energy gap-dependent effective temperature, showing that the effective temperature can be higher than the ambient bath temperature in the small energy gap limit and lower in the high bath temperature limit.
We then investigate how these motion-induced effective temperatures affect the operating conditions and performance of the quantum Otto cycle as a heat engine or a refrigerator. We show that the engine efficiency and refrigerator performance coefficient can exceed their standard Carnot bounds based on the bath's rest-frame temperatures, while remaining bounded by generalised Carnot limits expressed in terms of the effective temperatures experienced by the detector. In the high-temperature limit, the effective temperature coincides with that of an inertial detector moving through a thermal bath, and a comparison with the small-gap limit reveals that circular motion allows the standard Carnot bounds to be exceeded at lower speeds than in the inertial case.
\end{abstract}


\section{Introduction}

The Unruh effect \cite{Fulling:1972md, Davies:1974th, Unruh} states that a linearly uniformly accelerated observer with proper acceleration $a$ perceives the Minkowski vacuum of a relativistic quantum field as a thermal state at the Unruh temperature,
\begin{equation}
    T_U=\frac{\hbar a}{2\pi ck_B}.
\end{equation}
This temperature has an operational meaning in terms of the detailed balance condition satisfied by the excitation and de-excitation rates of a quantum detector carried by the observer, such as a qubit. 
While in general the motion of an observer will have an imprint in their detector's response, it is especially interesting to note that stationary detectors in Minkowski spacetime \cite{Letaw:1980yv} are driven to equilibrium states when they interact with a field in the Minkowski vacuum or more general 
steady states, in a sense that is made precise in \cite{BJA:DM}; see also \cite{Good_2020}.  
A prominent example is
circular motion, which, unlike linear acceleration, can be sustained within a laboratory of finite size for an arbitrarily long time.
In this case, the detector's response is not generally thermal at the Unruh temperature. Instead, an effective temperature can be defined operationally through the ratio of its excitation and de-excitation rates.  This temperature depends on the detector's energy gap and on the parameters of the circular trajectory, rather than on the proper acceleration alone \cite{Unruh:1998gq,Good_2020,Biermann:2020bjh}. 
Detectors in uniform circular motion interacting with fields in a thermal state have been studied previously in \cite{Hodgkinson:2014iua, Bunney:2023vyj,
Bunney:2023ipk}.

The thermal response of uniformly accelerated observers underlies the Unruh quantum Otto engine~\cite{arias2018unruh}. A standard quantum Otto cycle \cite{Abah:review} consists of two adiabatic strokes, during which the energy level spacing of the working medium is varied, and two isochoric strokes, during which the working medium exchanges heat with hot and cold reservoirs at temperatures $T_h$ and $T_c<T_h$, respectively. In the Unruh  Otto engine, the working medium is a qubit detector weakly coupled to a  scalar field. During the isochoric strokes, the detector undergoes uniform acceleration while interacting with the field in the Minkowski vacuum state, exhibiting a thermal response at the Unruh temperature. Alternating between two accelerations $a_h>a_c$ thus mimics interactions with hot and cold thermal reservoirs, so that positive work can be extracted under suitable conditions. This framework has been extended to various settings, including entangled qubits \cite{UOtto:entangl}, fermionic fields \cite{gray2018}, instantaneous detector-field interactions \cite{NK:DM,gallock2024relativistic}, and other accelerated trajectories, such as circular motion in the Minkowski vacuum \cite{gallock2023quantum,Sarkar:2025ooc}. More broadly, the influence of curved spacetime backgrounds, including black hole spacetimes, on the performance of relativistic thermal machines has also been investigated \cite{DEB20,Ferketic:Deffner,DM:BTZ}.

A detector moving at constant velocity relative to a thermal bath exhibits a non-thermal response due to Doppler shifts, characterised by energy gap-dependent effective temperatures that can be hotter or colder than the bath's temperature \cite{Costa:1994yx, Costa:1995yv, Hodgkinson:2014iua,  Papadatos:2020pbv, Passegger:2025amw}. Recently, the implications of this behaviour were investigated for a quantum Otto cycle whose working medium is a qubit detector undergoing inertial motion during the isochoric strokes \cite{DM:OA:Efficiency, DM:OA:TUR}, where generalised Carnot bounds on the efficiency of such a machine were derived in terms of the effective temperatures experienced by the detector. These bounds can exceed the standard Carnot bound expressed in terms of the bath rest-frame temperatures, so that higher efficiencies become possible without violating the second law of thermodynamics. 

In the present work, we extend these investigations to a quantum Otto thermal machine whose working medium is a qubit detector undergoing uniform circular motion in $(3+1)$-dimensional Minkowski spacetime while interacting with a scalar field in a thermal state during the isochoric strokes. Specifically, we consider a configuration in which the detector remains at rest with respect to one thermal bath and undergoes uniform circular motion relative to the other during the respective isochoric strokes. Unlike inertial motion, circular motion introduces non-zero proper acceleration while preserving the stationarity of the detector's response. We  calculate the response of the detector in circular motion and determine the energy gap-dependent effective temperature from the ratio of its excitation and de-excitation rates, identifying conditions under which it is higher or lower than the ambient bath temperature in the small energy gap and high bath temperature limits.  We then investigate how these motion-induced effective temperatures affect the conditions under which the Otto cycle operates as a heat engine or as a refrigerator, as well as its performance in each regime. We show that, as in the inertial case, both the engine efficiency and refrigerator coefficient of performance can exceed the standard Carnot bounds based on the bath rest-frame temperatures, while respecting generalised Carnot bounds expressed in terms of the effective temperatures experienced by the detector. In the high bath temperature limit, we find that the effective temperature coincides with that of an inertial detector moving through a thermal bath. Comparing this result with that obtained for circular motion in the small energy gap limit, we find that the latter can enable such enhancements at lower velocities.

The rest of the paper is organised as follows. In Section \ref{sec:model-and-eff-temperature} we introduce the Unruh-DeWitt detector, which serves as the qubit working medium, and define the effective temperature that plays a central role in the Otto cycle. Since the detector will interact with a scalar field in a thermal state, we also introduce the thermal Wightman function of the field towards the end of the section. In Section \ref{sec:quantum-otto} we review the quantum Otto cycle with the Unruh-DeWitt detector as the working medium, operating either as
a heat engine or as a refrigerator, together with the corresponding efficiency and coefficient of performance.
In Section \ref{Sec:circular} we study the response of a circularly moving detector coupled to a thermal bath
with particular attention to the small gap limit of the detector ($E \to 0$) and the high temperature limit of the field ($T \to \infty$). In Section~\ref{sec:otto-with-circular} we use these results to analyse the Otto engine and refrigerator with circularly moving
detectors, and show that their performance can exceed the standard Carnot bound defined by the bath rest-frame temperatures. Our closing remarks appear in Section \ref{sec:closing-remarks}. Some technical steps of the analysis are deferred to the Appendix.

We use units in which the speed of light, reduced Planck constant $\hbar$ and Boltzmann's constant $k_B$ are set to unity.  
In asymptotic formulae, 
$f(x)=O(g(x))$ denotes that $f(x)/g(x)$ is bounded in the limit of interest, and $f(x) = o(g(x))$ denotes that $f(x)/g(x)\to0$ in the limit of interest. 
The Fourier transform is defined as
\begin{equation}
    \widehat{f}(\omega) = \intinf \dd t \, \ee^{-i\omega t} f(t), 
    \label{eq:fourier-def}
\end{equation}
and $\sinc$ and $\sinch$ denote the analytic functions defined by 
\begin{subequations}
\begin{align}
\sinc z & =\begin{cases} \frac{\sin z}{z} & z\neq 0,\\ 1 & z=0,\end{cases}
\\
\sinch z & =\begin{cases} \frac{\sinh z}{z} & z\neq 0,\\ 1 & z=0 . \end{cases}
\end{align}
\end{subequations}

\section{Effective temperature measured by an Unruh-DeWitt detector}
\label{sec:model-and-eff-temperature}

In this section we briefly summarise our field-detector model and the emergence of a late time effective temperature in this model.

\subsection{Field-detector model}
\label{subsec:FieldDec}
We consider a pointlike Unruh-DeWitt (UDW) detector \cite{Unruh,DeWitt} on a prescribed worldline~$\x(\tau)$, parametrised by proper time~$\tau$, in $(3+1)$-dimensional Minkowski spacetime. We model the detector as a two-level quantum system (qubit) with 
Hilbert space $\mathcal{H}_D=\mathbb{C}^2$, 
the orthonormal basis vectors $\ket{g} = \left(\begin{smallmatrix}0\\1\end{smallmatrix}\right)$ and $\ket{e} = \left(\begin{smallmatrix}1\\0\end{smallmatrix}\right)$, and the free Hamiltonian $H_D=\frac12 E\sigma_z$, 
where $\sigma_z = \diag(1,-1)$
is the Pauli $z$-matrix and $E\in\mathbb{R}$. The eigenenergies of $\ket{g}$ and $\ket{e}$ are respectively $-\frac12 E$ and~$\frac12 E$. When $E>0$, $\ket{g}$ is the ground state and $\ket{e}$ is the excited state; when $E<0$, the roles of $\ket{g}$ and $\ket{e}$ are reversed. We denote the identity operator in $\mathcal{H}_D$ by~$I_D$.

In the spacetime, we consider 
a real massless Klein-Gordon field~$\phi$.
The field, viewed as an abstract algebra generator, obeys the Klein-Gordon equation as a distribution, $\Box \phi = (-\partial_t^2 + \Delta) \phi = 0$, and is subject to the commutation relations $[\phi(f), \phi(g)] = - i E(f,g)I_\phi$, where $f$ and $g$ are suitable test functions, 
$E(\cdot,\cdot)$ is the (advanced-minus-retarded) causal propagator of the differential operator $\Box$ and $I_\phi$ denotes the identity operator. 

We work in the interaction picture and take the total Hilbert space of the theory to be $\mathcal{H}_D \otimes \mathcal{H}_\phi$. We further suppose that the theory is describable by a total Hamiltonian of the form $H = I_D \otimes H_\phi + H_D \otimes 1_\phi + H_{\rm int}$, and thus that the initial state of the quantum field is stationary.
The interaction Hamiltonian is
\begin{align}
    H_{\text{int}}(\tau)=c\chi(\tau) \mu(\tau)\phi(\mathsf{x}(\tau)),
\label{eq:Hint-def}
\end{align}
where $c>0$ is a coupling constant, $\chi(\tau)$ is a real-valued switching function that specifies how the interaction is turned on and off, $\mu(\tau)= e^{i E \tau}\sigma_+ + e^{-i E \tau}\sigma_-$ is the detector's monopole moment operator expressed in
terms of the ladder operators $\sigma_+=\ket{e}\bra{g}$ and $\sigma_-=\ket{g}\bra{e}$, and $\phi(\mathsf{x}(\tau))$ is the pullback of $\phi$ to the detector’s worldline. 

 Although in the upcoming sections we will not explicitly deal with the Klein-Gordon Hamiltonian, $H_\phi$, we remark that in the thermal bath situation described in Section \ref{sec: thermal states} this operator can be written explicitly using the Araki-Woods representation \cite{Araki-Woods} (see also e.g.\ Chapter 17 of~\cite{Derezinski-Gerard}.)

Two quantities of prime interest for this model are the detector's transition probability and the detector's reduced final state upon tracing out the field degrees of freedom. We discuss these respectively in Section \ref{subsec: response-and-stationary} and Section~\ref{subsec:effective-temperature}. 

\subsection{Transition probability and stationary transition rate} \label{subsec: response-and-stationary}

We assume that the detector is initially prepared in the state $|g\rangle$ and the field is initially prepared in some Hadamard state with Wightman two-point function $G_+(\x, \x')$. 
After the interaction has ceased, the probability to find the detector in the state~$|e\rangle$, regardless of the final state of the field, can be found by time-dependent perturbation theory in the coupling constant~$c$, with the leading order outcome
\begin{align}
    \mathcal{P}(E) = c^2 | \langle e | \mu(0) | g\rangle|^2 \mathcal{F}_\chi(E) + O(c^3),
    \label{TransProb}
\end{align}
where $\mathcal{F}_\chi(E)$ is the response function, given by 
\begin{align} \label{eq: response function}
    \mathcal{F}_\chi(E) := \intinf \dd \tau \intinf \dd \tau' \, \chi(\tau)\chi(\tau')  \, \ee^{-i E (\tau - \tau')} \W(\tau,\tau'), 
\end{align}
and $\W(\tau, \tau') = G_+(\x(\tau), \x(\tau'))$ is the pullback of the Wightman function to the detector's worldline. 
To the leading order in~$c$, 
all the information about the detector's trajectory, the energy gap $E$ and the initial state of the field is hence encoded in $\mathcal{F}_\chi(E)$~\eqref{eq: response function}.

We now specialise to a detector following a stationary worldline, that is, an integral curve of a Killing vector that is timelike at least in a neighbourhood of the worldline~\cite{Letaw:1980yv}, and we further assume that the initial state of the field is invariant under the Poincar\'e transformations generated by this Killing vector. From the stationarity it follows that the squared spacetime interval between any two points on the trajectory depends only on the proper time difference of the points, $(\x(\tau)-\x(\tau'))^2 = (\x(\tau-\tau')-\x(0))^2$, and from the invariance of the field's initial state it further follows that the Wightman function pulled back to the worldline satisfies $\W(\tau,\tau') = \W(\tau-\tau',0)$. 
Defining $\W(s) := \W(s,0)$, we then have \cite{Waiting}
\begin{equation} \label{eq: stationary response}
    \F_\chi(E) = \frac{1}{2\pi}\intinf \dd \omega \, |\widehat{\chi}(\omega)|^2 \widehat{\W}(E+\omega), 
\end{equation}
under mild regularity assumptions on $\chi$ and~$\W$. 

We consider the limit in which the field and the detector interact for a long time. To implement this limit, we introduce a one-parameter family of switching functions $\chi_\lambda$, with $\lambda>0$ controlling the interaction duration. The family is chosen so that, as $\lambda$ grows, the support of the interaction is approximately contained on an interval of length proportional to~$\lambda$, while the  magnitude of the interaction remains approximately constant in an averaged sense over this interval. One then expects from \eqref{eq: stationary response} that $\F_{\chi_\lambda}$ grows linearly with $\lambda$, and hence that the relevant quantity is the rescaled response $\lambda^{-1}\F_{\chi_\lambda}(E)$, which has a finite limit as $\lambda \to \infty$. For suitable switching families, this expectation is realised, with the outcome that
\begin{equation} \label{eq: long time limit}
    \lambda^{-1}\F_{\chi_\lambda}(E) \xrightarrow[\lambda \to \infty]{}\alpha \widehat{\W}(E),
\end{equation}
for some constant $\alpha>0$, where the limit is understood pointwise in $E$ \cite{Waiting, Parry:2025wub}. A~sufficient set of conditions for \eqref{eq: long time limit} to hold was given in the asymptotically scaled switching family formalism of \cite{Parry:2025wub, Parry:2026ggh}; the main result is that, given such a family $\chi_\lambda$, there exists a square-integrable function $\xi \neq 0$ such that $\widehat{\xi}(u) = \lim_{\lambda \to \infty}[\lambda^{-1}\widehat{\chi}_\lambda(u/\lambda)]$, and the limit \eqref{eq: long time limit} holds with $\alpha = \Vert \xi \Vert^2$.

The long time limit described above is consistent with the usual informal prescription for long interactions, in which one factors out the total interaction time from the response function and interprets the remaining quantity as the transition rate \cite{Unruh, DeWitt, Birrell:1982ix}. In what follows, we take $\alpha=1$ in \eqref{eq: long time limit} and refer to $\widehat{\W}(E)$ as the detector transition rate in the long time limit.

\subsection{Detector's reduced final state and effective temperature}
\label{subsec:effective-temperature}

We now use the transition rate $\widehat{\W}(E)$ to introduce an effective temperature that characterises the detector's response in the long time limit implementations in which \eqref{eq: long time limit} holds. 

We first recall a standard result in the open quantum systems framework. Consider a detector following a stationary worldline, with the field prepared in a state that is invariant under the corresponding one-parameter group of Poincar\'e transformations. In the Born-Markov ansatz, the field is modelled as a Markovian environment and the field-detector coupling constant $c$ is taken to be sufficiently small that the total state of the combined system remains approximately uncorrelated over the relevant timescales. 
Using the adiabatic switching family $\chi_\lambda(\tau)=\chi(\tau/\lambda)$ to implement the long time limit, it is shown in \cite{DeBievre:2006pys, BJA:DM}
that the detector's reduced density matrix approaches at late times the stationary state
\begin{equation} \label{GibbsEff}
    \rho(E) = \frac{\exp\!\left( -H_D/T^\eff(E) \right)}{\tr\!\left[\exp\!\left( -H_D/T^\eff(E) \right) \right]} =  \frac{1}{1 + \exp\!\big(-E/T^\eff(E)\big)} \begin{pmatrix} \exp\!\big(-E/T^\eff(E)\big) & 0 \\ 0 & 1 \end{pmatrix},
\end{equation}
in the matrix notation for the basis 
$\left\{ \ket{g}, \ket{e}\right\}$ introduced in Section~\ref{subsec:FieldDec},
where 
\begin{equation} \label{eq: effective temp}
     T^\eff(E) = \frac{E}{\log\!\left(\displaystyle{\frac{\widehat{\W}(-E)}{\widehat{\W}(E)}} \right)},
\end{equation}
whenever the right-hand side of \eqref{eq: effective temp} is defined. 

Note that \eqref{GibbsEff} and \eqref{eq: effective temp} hold both for $E>0$ and $E<0$, and $T^\eff(E) = T^\eff(-E)$, so that $T^\eff(E)$ depends on $E$ only through~$|E|$. The distinction between positive and negative values of $E$ in the stationary state \eqref{GibbsEff} is purely notational, recalling from Section \ref{subsec:FieldDec} that the sign of $E$ encodes which of the basis vectors $\left\{\ket{g},\ket{e}\right\}$ is the ground state and which is the excited state: in either case, the expectation value of the excited state is suppressed by the factor $\exp\!\big(-|E|/T^\eff(E)\big)$.

When $T^\eff(E)$ is independent of $E$, $\rho(E)$ \eqref{GibbsEff} is a thermal Gibbs state and $T^\eff(E)$ is the corresponding detailed balance temperature \cite{Einstein,terhaar-book}. For example, this occurs for a uniformly linearly accelerated detector in the Minkowski vacuum, where $T(E)=a/(2\pi)$ for proper acceleration $a>0$; this is the Unruh effect \cite{Fulling:1972md, Davies:1974th, Unruh, Fulling:2014}. In more general stationary settings, $T^\eff(E)$ generally depends on $E$, and therefore $\rho(E)$ \eqref{GibbsEff} is not strictly thermal. However, whenever the dependence of $T^\eff(E)$ on $E$ is sufficiently weak on a given interval, $\rho(E)$ may still be viewed as approximately thermal at the gap-dependent effective temperature $T^\eff(E)$. For further studies of $T^\eff(E)$ in stationary detector settings, see \cite{Biermann:2020bjh, Good_2020, Bunney:2023vyj, Bunney:2023ipk, Parry:2024jrm, Passegger:2025amw, Parry:2025wub}.

We shall use the final reduced state $\rho(E)$ \eqref{GibbsEff} in the Otto cycle analysis in Section~\ref{sec:otto-with-circular}.

\subsection{Stationary detectors in a thermal bath} \label{sec: thermal states}
In the applications in this paper, we shall be interested in the case in which the field is prepared in a thermal state, and here we briefly identify the stationary detector trajectories that will be relevant in the analysis below.

In inertial coordinates $\x=(t,\mathbf{x})$, consider a thermal state at temperature $T>0$ with respect to the time translations generated by $\partial_t$. The Lorentz frame defined by these coordinates is the rest frame of the corresponding thermal bath. For a massless scalar field, the Wightman function in this thermal state is
\cite{Dowker:1978aza}
\begin{equation} \label{eq: thermal Wightman}
    \W_T(\x,\x') = -\frac{T}{4\pi |\mathbf{x}-\mathbf{x'}|}\frac{\sinh\!\left(2\pi T|\mathbf{x}-\mathbf{x'}| \right)}{\cosh\!\left(2\pi T(t-t'-i\epsilon) \right) - \cosh\!\left(2\pi T|\mathbf{x}-\mathbf{x'}|\right)},
\end{equation}
where the distributional $\epsilon\to 0^+$ limit is understood, and we have introduced the subscript $T$ to indicate the ambient temperature. 

The thermal Wightman function \eqref{eq: thermal Wightman} is invariant under temporal translations, spatial translations and spatial rotations in the bath rest frame. It follows that for a detector moving inertially at constant speed with respect to the bath, the pullback of $\W_T$ to the detector's worldline depends only on the proper time difference, since the worldline is generated by a combination of temporal and spatial translations. Similarly, the same is true for a detector in uniform circular motion about a centre at rest in the bath frame, since the worldline is generated by a combination of a temporal translation and a spatial rotation. 

For both inertial motion and uniform circular motion in a thermal bath, the results of Sections \ref{subsec: response-and-stationary} and \ref{subsec:effective-temperature} apply, and in particular, at late times the detector's transition rate is given by $\widehat{\W}_T(E)$ and the detector's state approaches the stationary state \eqref{GibbsEff} at the effective temperature $T^\eff(E)$ \eqref{eq: effective temp}. For a moving detector, $T^\eff(E)$ generally differs from the temperature of the thermal bath, and in particular it can be either hotter or colder than the ambient temperature; see \cite{Bunney:2023vyj, DM:OA:Efficiency} for example. In the special case of a static worldline, the detector genuinely thermalises at late times at the ambient bath temperature.

Both the static and uniform circular trajectories (see \eqref{eq: CM worldline}) will be relevant to the protocols of Sections \ref{sec:quantum-otto} and \ref{sec:otto-with-circular}.

\section{Quantum Otto thermal machines}
\label{sec:quantum-otto}

In this section, we describe the quantum Otto cycle that uses a two-level UDW detector as its working medium. Then, we present the conditions for this cycle to operate as a heat engine and then a refrigerator.

\subsection{Quantum Otto cycle}
Consider a finite-dimensional quantum system in a given state $\rho$ with Hamiltonian $H$ and mean internal energy $\expval{U}=\tr[\rho H]$.  During a thermodynamic process, changes in the state and/or the Hamiltonian generically lead to a change in the mean internal energy $\Delta\!\expval{U}$. This change can be decomposed into contributions from  work $\expval{W}$ and heat $\expval{Q}$, defined, respectively, as \cite{Alicki_1979}
\begin{subequations} \label{eq: heat and work defs}
    \begin{align}
        \expval{W} &= \int \dd \tau \, \tr\!\Big[\rho(\tau) \, \dot{H}(\tau)
        \Big], \label{eq: work def} \\
        \expval{Q} &= \int \dd 
        \tau \, \tr\!\Big[H(\tau) \,  \dot{\rho}(\tau)
        \Big]. \label{eq: heat def}
    \end{align}
\end{subequations}
Then, the first law of thermodynamics reads $\Delta \!\expval{U} = \expval{W} + \expval{Q}$.
With these definitions, positive (negative) work done $\expval{W}$ corresponds to work done on (extracted from) the system, while positive (negative) heat  $\expval{Q}$ corresponds to heat absorbed (released) by the system. 
Here, we describe the quantum Otto cycle, using the two-level UDW detector described in Section \ref{subsec:FieldDec} as its working medium. For notational convenience, we assume throughout Section \ref{sec:quantum-otto} that the detector has $E>0$, so that $\ket{g}$ and $\ket{e}$ are respectively the ground state and the excited state.

\begin{figure}[t!]
    \centering
    \includegraphics[width=0.5\linewidth]{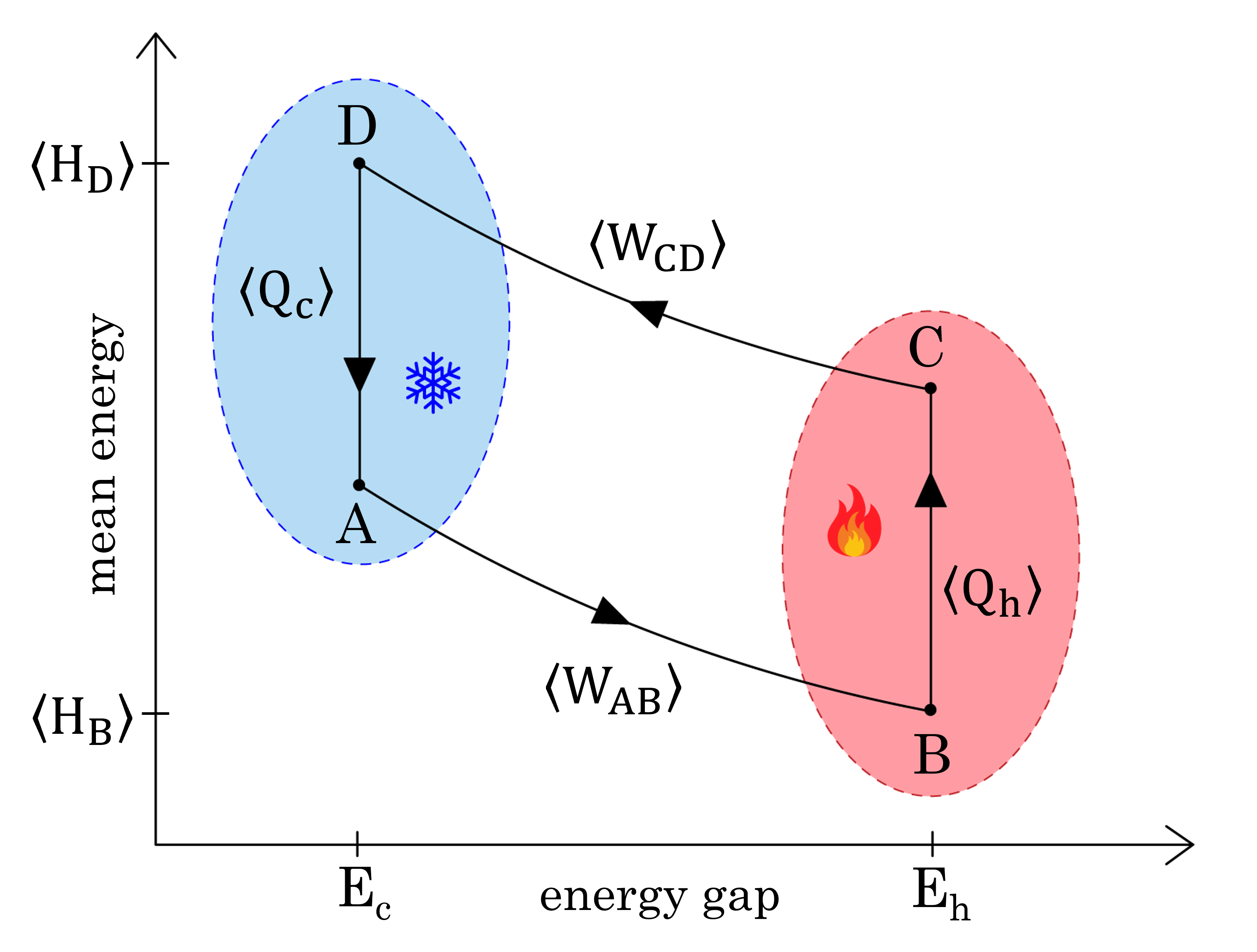}
    \caption{The quantum Otto cycle consists of four strokes: two adiabatic strokes, during which the energy gap of the qubit working medium is varied, with no coupling to the ambient field, intercalated with two isochoric strokes, during which the working medium is coupled to thermal baths, a hot bath at temperature $T_h$ and a cold bath at temperature $T_c<T_h$. During the isochoric strokes, the qubit may follow a circular trajectory and experience frequency-dependent effective temperatures that are either hotter or colder than the corresponding ambient bath temperatures.}
    \label{fig:Otto}
\end{figure}

The cycle operates between two thermal field baths, a cold bath at temperature $T_c$, and a hot bath at temperature $T_h>T_c$, and consists of four consecutive strokes; see Figure \ref{fig:Otto}. These comprise two adiabatic strokes during which the detector is isolated from the field and its energy gap is varied, alternating with two isochoric strokes during which the energy gap is held fixed, and the detector is coupled to one of the two heat baths. 

We focus on a particular implementation of the Otto cycle in which, during the isochores, the detector is static relative to one thermal bath and in uniform circular motion relative to the other. As discussed in Section \ref{sec: thermal states}, both types of motion give a stationary response. When coupled to the thermal baths, we assume that each isochore lasts sufficiently long for the detector to reach the stationary state \eqref{GibbsEff}
\begin{equation} \label{GibbsEff2}
    \rho_i(E) = \frac{1}{1 + \exp\!\big(-E/T_i^\eff(E)\big)} \begin{pmatrix} \exp\!\big(-E/T_i^\eff(E)\big) & 0 \\ 0 & 1 \end{pmatrix},
\end{equation}
where now the subscript $i=c,h$ indicates the cold and hot thermal baths, respectively. Note that for the static detector, $T_i^\eff(E)=T_i$ coincides with the bath temperature. 

We now describe the four strokes in turn and evaluate the work and heat exchanged during each stroke using \eqref{eq: work def} and~\eqref{eq: heat def}. We denote the four corners of the cycle by $A,B,C,D$, with the cycle proceeding in the order $A\to B \to C \to D \to A$, as is depicted in Figure~\ref{fig:Otto}. The detector is prepared initially in the state~$\rho_c(E_c)$, and the cycle proceeds with the following four steps:

\begin{itemize}
    \item[\textbf{1.}] \textbf{Adiabatic expansion} $(A\to B)$: The detector is isolated from the thermal bath and undergoes an adiabatic expansion, during which its energy gap increases from $E_c$ to $E_h$, with $E_h>E_c>0$. The detector's state evolves unitarily under the Hamiltonian $H_D(\tau) = \frac{1}{2}E(\tau)\sigma_z$. Then since $H_D(\tau)$ commutes with the detector's initial state $\rho_c(E_c)$ throughout this stroke, it follows that $\dot{\rho}(\tau) = 0$, and thus by \eqref{eq: heat def}, no heat is exchanged during the process. The corresponding change in internal energy is thus entirely due to work $\expval{W_{AB}}$. The formula \eqref{eq: work def} then gives
    \begin{equation}\label{eq: work AB}
     \expval{W_{AB}} = \frac{E_c-E_h}{2}\tanh\!\left(\frac{E_c}{2T^\eff_c(E_c)} \right).
    \end{equation}
    Since $E_c<E_h$, we have $\expval{W_{AB}}<0$, so this work is extracted from the detector.

    \item[\textbf{2.}] \textbf{Hot isochore} $(B\to C)$: The detector is then placed in contact with the hot thermal bath at temperature $T_h$, while keeping the energy gap $E_h$ fixed. The detector follows either a static or uniform circular trajectory and at late times relaxes to the stationary state $\rho_h(E_h)$, characterised by the effective temperature $T^\eff_h(E_h)$. During this stroke, $H_D = \frac{1}{2}E\sigma_z$ for all times, so that $\dot{H}=0$, and thus, by \eqref{eq: work def}, no work is done during this process. The change in internal energy is therefore entirely due to heat $\expval{Q_{BC}}$. The formula \eqref{eq: heat def} then gives
    \begin{equation} \label{eq: heat BC}
        \expval{Q_{BC}}\equiv\expval{Q_{h}} = \frac{E_h}{2}\left( \tanh\!\left(\frac{E_c}{2T^\eff_c(E_c)} \right) - \tanh\!\left(\frac{E_h}{2T^\eff_h(E_h)} \right)\right).
    \end{equation}
\item[\textbf{3.}] \textbf{Adiabatic compression} $(C \to D)$: The detector is again isolated from the thermal bath and undergoes an adiabatic compression, during which its energy gap decreases from $E_h$ back to its initial value $E_c$. As in step 1, no heat is exchanged and the change in internal energy is due to work $\expval{W_{CD}}$, given by
    \begin{equation}
        \expval{W_{CD}} = \frac{E_h-E_c}{2}\tanh\!\left(\frac{E_h}{2T^\eff_h(E_h)} \right).
    \end{equation}
    Since $E_c<E_h$, we have $\expval{W_{CD}}>0$, so this work is done on the detector.

    \item[\textbf{4.}] \textbf{Cold isochore} $(D\to A)$: The detector is then placed in contact with the cold thermal bath at temperature $T_c$, while keeping the energy gap $E_c$ fixed. The detector again follows either a static or uniform circular trajectory, and at late time relaxes to the stationary state $\rho_c(E_c)$, characterised by the effective temperature $T^\eff_c(E_c)$. As in step 2, no work is done during this stroke, and the change in internal energy is entirely due to heat $\expval{Q_{DA}}$, given by
    \begin{equation} \label{eq: heat DA}
        \expval{Q_{DA}}\equiv\expval{Q_{c}} = \frac{E_c}{2}\left( \tanh\!\left(\frac{E_h}{2T^\eff_h(E_h)} \right) - \tanh\!\left(\frac{E_c}{2T^\eff_c(E_c)} \right)\right).
    \end{equation}
    
\end{itemize}

We note that, in practice, there are energy costs associated with accelerating the detector in preparation for steps 2 and 4, and with moving the detector between the two thermal baths. 
Here, we assume, as is standard, that the detector's motion is externally sustained and its trajectory is prescribed, without explicitly modelling the associated energetic cost or backreaction \cite{Hu:2012jr}. This is analogous to standard treatments in which the energetic cost of engineering nonequilibrium reservoirs is neglected \cite{Huang:2012,OA:2014,Long:2015,Hu_sqOtto}. 

Overall, the net work associated with the Otto cycle is
\begin{equation}\label{eq: net work}
     \Delta \! \expval{W} = \expval{W_{AB}} + \expval{W_{CD}} = \frac{E_h-E_c}{2}\!\left(\tanh\!\left(\frac{E_h}{2T^\eff_h(E_h)} \right)- \tanh\!\left(\frac{E_c}{2T^\eff_c(E_c)} \right) \right),
\end{equation}
and when $\Delta \! \expval{W}<0$ this net work is extracted from the system, while when $\Delta \! \expval{W}>0$, this net work is done on the system. We now use the sign of $\Delta \! \expval{W}$ to distinguish between two modes of operation of the Otto cycle; as a heat engine and as a refrigerator.

\subsection{Quantum Otto heat engine} \label{sec: Otto heat engine}

\subsubsection{Heat engine condition}
The Otto cycle described above operates as a heat engine when it absorbs heat from the hot bath during the hot isochore $BC$, releases heat to the cold bath during the cold isochore $DA$, and net work is extracted;
\begin{equation} \label{eq: heat engine conditions}
    \expval{Q_h}>0,\qquad\quad \expval{Q_c}<0, \qquad\quad \Delta \!\expval{W}<0,
\end{equation}
where recall from \eqref{eq: heat BC} and \eqref{eq: heat DA} that $\expval{Q_h}:= \expval{Q_{BC}}$ and $\expval{Q_c}:=\expval{Q_{DA}}$. 

From \eqref{eq: heat BC}, \eqref{eq: heat DA} and \eqref{eq: net work}, one finds that the inequalities \eqref{eq: heat engine conditions} are satisfied if and only if
\begin{equation} \label{eq: net work extraction condition}
     \frac{T^\eff_c(E_c)}{T^\eff_h(E_h)}< \frac{E_c}{E_h} < 1,
\end{equation}
where the inequality $E_c/E_h<1$ holds by assumption.
The condition \eqref{eq: net work extraction condition} is required for net work extraction from the Otto cycle, and it generalises the condition, $E_c/E_h > T_c/T_h$, by replacing the bath temperatures with the relevant effective temperatures.

We note that it is possible in our protocol to find parameter regimes in which $T^\eff_c(E_c) > T^\eff_h(E_h)$, however this would be inconsistent with the the heat engine condition \eqref{eq: net work extraction condition}. We shall therefore restrict $T^\eff_c(E_c) < T^\eff_h(E_h)$.

\subsubsection{Efficiency}
Denoting the total work output of the engine by $\expval{W_\mathrm{out}} := - \Delta\! \expval{W}>0$, the efficiency $\eta$ of the heat engine is defined by
\begin{equation} \label{eq: efficiency}
    \eta = \frac{\expval{W_\mathrm{out}}}{\expval{Q_h}},
\end{equation}
the ratio of the total work produced per cycle to the heat extracted from the hot thermal bath during the hot isochore $BC$. From \eqref{eq: heat BC} and \eqref{eq: net work}, we have $\expval{W_\mathrm{out}} = (1-\frac{E_c}{E_h})\expval{Q_{h}}$, and hence the efficiency \eqref{eq: efficiency} evaluates to
\begin{equation} \label{eq: efficiency 2}
    \eta = 1-\frac{E_c}{E_h}.
\end{equation}

The maximum efficiency of the heat engine can be derived from the second law of thermodynamics and the positivity of the entropy production. The entropy production in one cycle is given by \cite{abah2014efficiency}
\begin{equation} \label{eq: entropy production}
    \Sigma = S(\rho_B \Vert \rho_C) + S(\rho_D \Vert \rho_A) \geq 0,
\end{equation}
where $S(\rho\Vert \sigma):= \tr\!\left[\rho\ln \rho -  \rho \ln \sigma\right]$ is the quantum relative entropy of $\rho$ with respect to $\sigma$, and $\rho_A$, $\rho_B$, $\rho_C$ and $\rho_D$ are the states of the detector at the vertices of the Otto cycle, depicted in Figure \ref{fig:Otto}. In our case, $\rho_A=\rho_B = \rho_c(E_c)$ and $\rho_C=\rho_D=\rho_h(E_h)$, and an elementary calculation shows that 
\begin{equation} \label{eq: entropy production heat engine}
    \Sigma = \left( \frac{1}{T_c^\eff(E_c)} - \frac{1}{T_h^\eff(E_h)}\right) \expval{Q_{h}} - \frac{1}{T_c^\eff(E_c)}\expval{ W_\mathrm{out}}.
\end{equation}
Therefore, by combining the non-negativity of $\Sigma$ with \eqref{eq: efficiency 2} and \eqref{eq: entropy production heat engine}, one finds
\begin{equation} \label{eq: max efficiency}
    \eta \leq 1- \frac{T^\eff_c(E_c)}{T^\eff_h(E_h)}:=\eta_C^\eff,
\end{equation}
with equality when the engine operates with maximal efficiency, given by the generalised Carnot efficiency $\eta_C^\eff$\cite{abah2014efficiency,DM:OA:Efficiency,DM:OA:TUR}.
Recall from \eqref{eq: net work extraction condition} that $T_c^\eff(E_c)/T_h^\eff(E_h)<1$ is required for the cycle to operate as a heat engine, hence $0<\eta_C^\eff<1$.

\subsubsection{Efficiency at maximum power} \label{sec: EMP}
A more useful measure of  heat engine performance than the maximum
efficiency, which is attained only in the limit of vanishing power output, is the \emph{efficiency at maximum power} (EMP) \cite{Curzon:Ahlborn}. The power output over one cycle is defined as 
\begin{equation} \label{eq: power}
    P = \frac{\expval{W_{\rm out}}}{\tau_{\rm cycle}},
\end{equation}
where $\tau_{\rm cycle}$ denotes the total duration of the cycle, which we take to be long but fixed. Although the stationary states during the isochoric steps are strictly reached only in the infinitely long time limit, equilibration is effectively achieved when the interaction duration is sufficiently long compared with the characteristic relaxation time \cite{BJA:DM,DM:BTZ}. 

The EMP is then computed by first maximising the power output $P$ with respect to the energy gap $E_h$, while keeping the remaining parameters fixed, including the cycle duration $\tau_\mathrm{cycle}$, and subsequently computing the corresponding engine efficiency using \eqref{eq: efficiency 2}. We denote the EMP of the heat engine by $\eta^*$. We emphasise that this optimization is performed at a prescribed cycle duration rather than over the thermalisation times themselves. For simplicity, the durations of the adiabatic strokes are neglected.

\subsection{Quantum Otto refrigerator}
\subsubsection{Refrigerator condition}

The quantum Otto cycle described above operates as a refrigerator when it extracts heat from the cold bath during the cold isochore DA, and releases heat to the hot bath during the hot isochore BC, requiring net work to be done;
\begin{equation} \label{eq: fridge conditions}
    \expval{Q_h}<0,\qquad\quad \expval{Q_c}>0, \qquad\quad \Delta \!\expval{W}>0,
\end{equation}
where we recall from \eqref{eq: heat BC} and \eqref{eq: heat DA} that $\expval{Q_h}:= \expval{Q_{BC}}$ and $\expval{Q_c}:=\expval{Q_{DA}}$.

In this case, using \eqref{eq: heat BC}, \eqref{eq: heat DA} and \eqref{eq: net work}, one finds that the inequalities \eqref{eq: fridge conditions} are satisfied if and only if
\begin{equation} \label{eq: net work input condition}
    \frac{E_c}{E_h} < \frac{T^\eff_c(E_c)}{T^\eff_h(E_h)}.
\end{equation}
The condition \eqref{eq: net work input condition} is required for net work done on the detector during the Otto cycle, and it again generalises the condition, $E_c/E_h < T_c/T_h$ by replacing the bath temperatures with the relevant effective temperatures. As in the heat engine case, we further restrict $T^\eff_c(E_c) < T^\eff_h(E_h) $.

\subsubsection{Coefficient of performance}
Denoting the total work done on the refrigerator by $\expval{W_\mathrm{on}} := \Delta \!\expval{W}$, the performance of the quantum Otto refrigerator can be quantified by the coefficient of performance (COP)
\begin{equation} \label{eq: COP}
    \varepsilon = \frac{\expval{Q_c}}{\expval{W_\mathrm{on}}},
\end{equation}
the ratio of the heat extracted from the cold bath during the cold isochore $DA$ to the total work input per cycle. From \eqref{eq: heat DA} and \eqref{eq: net work}, we have $\expval{W_\mathrm{on}} = \left(E_h/E_c-1 \right)\expval{Q_c}$, and hence the COP \eqref{eq: COP} evaluates to 
\begin{equation} \label{eq: COP 2}
    \varepsilon = \frac{1}{E_h/E_c-1}.
\end{equation}

Using $\expval{Q_h} = -\expval{W_\mathrm{on}} - \expval{Q_c}$, the entropy production per cycle $\Sigma$ \eqref{eq: entropy production heat engine} may be rewritten as
\begin{equation} \label{eq: entropy production fridge}
    \Sigma = \left( \frac{1}{T_h^\eff(E_h)} - \frac{1}{T_c^\eff(E_c)}\right) \expval{Q_c} + \frac{1}{T_h^\eff(E_h)}\expval{W_\mathrm{on}}.
\end{equation}
Then, by combining the non-negativity of $\Sigma$ with \eqref{eq: COP 2} and \eqref{eq: entropy production fridge}, one finds
\begin{equation} \label{eq: Carnot COP}
    \varepsilon \leq \frac{1}{T^\eff_h(E_h)/T^\eff_c(E_c)-1}:= \varepsilon_C^\eff,
\end{equation}
with equality when the refrigerator operates with maximal performance, given by the generalised Carnot COP $\varepsilon_C^\eff$. It follows from the requirement that $T^\eff_c(E_c)/ T^\eff_h(E_h)<1 $ that $\varepsilon^\eff_C >0$. Unlike the efficiency of the heat engine, the COP of the refrigerator is not bounded from above by unity and can, in principle, become arbitrarily large.

\subsubsection{COP at maximum figure of merit} \label{sec: COP maximum figure}
A useful measure of the performance of the refrigerator is the figure of merit $\chi_{\rm merit}$, which is defined as the product of the coefficient of performance and the cooling power \cite{Abah_2016_refr}
\begin{equation} \label{eq: figure of merit}
    \chi_{\rm merit} = \varepsilon \frac{\expval{Q_c}}{\tau_\mathrm{cycle}},
\end{equation}
where, as explained in Section \ref{sec: EMP}, we take $\tau_\mathrm{cycle}$ to be fixed but sufficiently long as compared with the characteristic relaxation time, so that effective thermalisation is achieved during the isochoric steps  \cite{BJA:DM,DM:BTZ}. One may then compute the \textit{COP at maximum figure of merit} by first maximising $\chi_{\rm merit}$ with respect to the energy gap $E_c$, while keeping the remaining parameters fixed, and subsequently computing the corresponding COP using \eqref{eq: COP 2}. We denote the COP at maximum figure of merit by $\varepsilon^*$.

\subsection{Remark}

The effective temperatures entering the generalised Carnot efficiency \eqref{eq: max efficiency} and COP \eqref{eq: Carnot COP} can be tuned by varying the parameters of the detector's trajectory as well as the detector's gap. These effective temperatures can be either higher or lower than the temperature of the ambient thermal bath, thereby allowing performance beyond the standard Carnot bounds \cite{Callen}, $\eta_C=1-T_c/T_h$, and $\varepsilon_C=T_c/(T_h-T_c)$ for a heat engine and a refrigerator, respectively, expressed in terms of the bath rest-frame temperatures. In the next section, we identify these regimes for a detector undergoing uniform circular motion.

\section{Circular motion in a $3+1$ thermal bath}\label{Sec:circular}

We now consider a detector in uniform circular motion in $(3+1)$-dimensional Minkowski spacetime, with the field prepared in a thermal state. We take the centre of the circular trajectory to be at rest with respect to the associated thermal bath, so that there is no overall drift through the bath. As we shall see, such a detector has a stationary response. We then compute the corresponding effective temperature, focusing on the limit of small detector gap and the limit of high ambient temperature. These limits will be relevant in the Otto cycle analysis in Section~\ref{sec:otto-with-circular}.

\subsection{Detector response}

We consider the UDW detector model of Section~\ref{sec:model-and-eff-temperature}. For notational convenience, we allow the energy gap $E$ to take both positive and negative values. 

We prepare the ambient field in the thermal state of temperature $T>0$ described in Section~\ref{sec: thermal states}. 
In the Lorentz frame $\x = (t,x,y,z)$ that coincides with the rest frame of the bath,
we take the detector to undergo uniform circular motion
\begin{align} \label{eq: CM worldline}
    \x(\tau) = \big(\gamma \tau, R \cos(\gamma \Omega \tau), R \sin(\gamma \Omega \tau), 0 \big),
\end{align}
where $R>0$, $0<\Omega < 1/R$, $\gamma=(1-\Omega^2 R^2)^{-1/2}$, and $\tau$ is the proper time. The spatial projection of this worldline is a circle of radius $R$ with angular speed $\Omega$ with respect to the bath rest frame, and the corresponding proper acceleration is $a=\gamma^2\Omega^2 R$.
This circular worldline is an integral curve of the helical Killing vector
\begin{equation} \label{eq: CM Killing}
    \xi(\x) = \gamma \partial_t - \gamma \Omega (y \partial_x - x \partial_y).
\end{equation}
$\xi$ is timelike in the region $x^2+y^2 < 1/\Omega^2$, and the worldline \eqref{eq: CM worldline} lies in this region by the assumption $0<\Omega < 1/R$. 

Since the thermal Wightman function \eqref{eq: thermal Wightman} is invariant under time translations and spatial rotations in the bath rest frame, it is invariant under the flow generated by $\xi$. Consequently, the pullback of $\W_T(\x,\x')$ to the circular worldline is stationary, depending on $\tau$ and $\tau'$ only through their difference $s=\tau-\tau'$.
Substituting \eqref{eq: CM worldline} into \eqref{eq: thermal Wightman}, and isolating the distributional singularity at $s=0$, one finds
\begin{equation} \label{eq: Wightman pullback}
    \W_T(s) =  -\frac{1}{4\pi^2(s-i \epsilon)^2} + \frac{1}{4\pi^2 s^2}  -\frac{T^2}{2}\frac{\sinch\!\left(4\pi T R\sin\!\left(\tfrac{\gamma \Omega}{2}s \right)\right)}{\cosh\!\left( 2\pi T \gamma s \right) - \cosh\!\left( 4\pi T R\sin\!\left(\tfrac{\gamma \Omega}{2}s \right) \right)},
\end{equation}
where recall that $\W_T(s):=\W_T(s,0)$.
The distributional behaviour of \eqref{eq: Wightman pullback} is contained entirely in the first term, while the sum of the second and third terms is regular everywhere, including $s=0$; note in particular that the denominator in the third term is positive for $s\ne0$, using $\Omega R<1$. The first term is precisely the Minkowski vacuum Wightman function evaluated on an inertial worldline, while the remaining regular contribution encodes both the dependence on the ambient temperature and on the detector's circular motion through the bath. 

As reviewed in Section~\ref{sec:model-and-eff-temperature}, the long time effective temperature is determined by the long time transition rate $\widehat{\W}_T(E)$. Taking the distributional Fourier transform of~\eqref{eq: Wightman pullback}, we may  split $\widehat{\W}_T(E)$ into the transition rate of an inertial detector in the vacuum and a correction as
\begin{subequations} \label{eq: CM thermal response}
    \begin{align}
      \widehat{\W}_T(E) &= \widehat{\W}^\mathrm{in}(E) + \widehat{\W}_T^\mathrm{corr}(E), \\
       \widehat{\W}^\mathrm{in}(E) &= -\frac{E }{2\pi}\Theta(-E),  \label{eq:Fin(E)}\\
       \widehat{\W}_T^\mathrm{corr}(E) &= \frac{T}{\pi \gamma} \intoinf \dd z \, \cos\!\left(\tfrac{2 \tilde E}{\gamma}z \right)\!\left( \frac{\gamma^2}{z^2} - \frac{\sinch\!\left(\tilde R\sin\!\left(\tilde \Omega z \right)\right)}{2\left[\cosh(z) - \cosh\!\left( \tilde R\sin\!\left(\tilde \Omega z \right) \right)\right]} \right), \label{eq: Fcorr}
    \end{align}
\end{subequations}
where in \eqref{eq: Fcorr} we have changed variables by $z=2\pi T \gamma s$, used the evenness of the expression in parentheses, and introduced the dimensionless parameters
\begin{equation} \label{eq: dimensionless variables}
    \tilde E = \frac{E}{4\pi T} , \qquad \tilde \Omega = \frac{\Omega}{4\pi T} , \qquad \tilde R = 4\pi T R.
\end{equation}
Note that in terms of these dimensionless parameters, the orbital speed is $v=\tilde \Omega \tilde R$.

We now compute the small gap and high ambient temperature limits of \eqref{eq: CM thermal response} and the associated effective temperature, using \eqref{eq: effective temp}.

\subsection{Small gap limit}\label{sec: CM small gap}
First, consider the $E\to 0$ limit of \eqref{eq: CM thermal response} with all other parameters held fixed. This limit was considered in \cite{Bunney:2023ipk}, where it was shown that \eqref{eq: CM thermal response} has the small $E$ expansion
\begin{equation}\label{eq: response small gap expansion}
    \widehat{\W}_T(E) = \frac{T}{\pi \gamma}\intoinf \dd z \,\left( \frac{\gamma^2}{z^2} - \frac{\sinch\!\left(\tilde R\sin\!\left(\frac{v}{\tilde R} z \right)\right)}{2\!\left[\cosh(z) - \cosh\!\left( \tilde R\sin\!\left(\frac{v}{\tilde R} z \right) \right)\right]} \right) - \frac{1}{4\pi}E + O(E^2).
\end{equation}
By substituting \eqref{eq: response small gap expansion} into \eqref{eq: effective temp}, one then finds that the ratio of the effective detailed balance temperature to the ambient thermal bath temperature, $T^\eff(E)/T$, in the small $E$ limit, is
\begin{equation} \label{eq:beta T(E) small E}
    \frac{T^\eff(E)}{T} = \frac{2}{\gamma} \intoinf \dd z \,\left( \frac{\gamma^2}{z^2} - \frac{\sinch\!\left(\tilde R\sin\!\left(\tilde \Omega z \right)\right)}{2\!\left[\cosh(z) - \cosh\!\left( \tilde R\sin\!\left(\tilde \Omega z \right) \right)\right]} \right) + O(E^2).
\end{equation}

\begin{figure}
    \centering
    \includegraphics[width=0.48\linewidth]{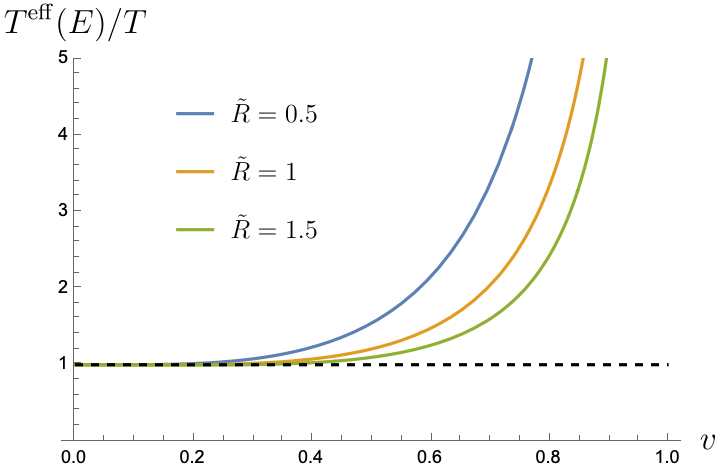}\hspace{0.4cm}   \includegraphics[width=0.48\linewidth]{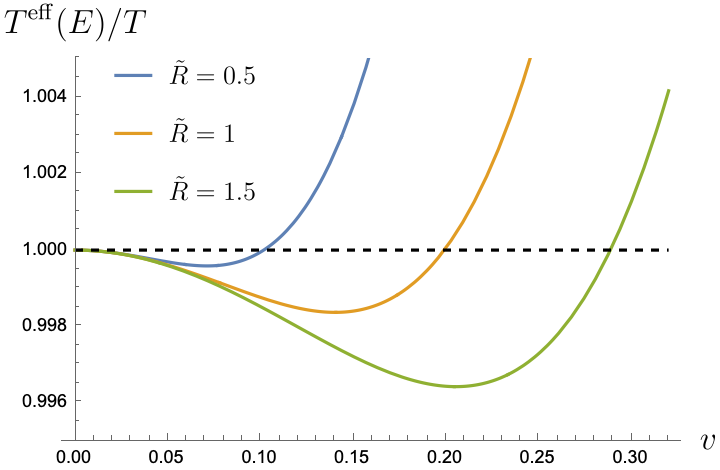} 
    \caption{Leading order term in the small $E$ expansion \eqref{eq:beta T(E) small E} of the temperature ratio, $ T^\eff(E)/T$, as a function of the orbital speed $v=\tilde \Omega \tilde R$. The curves correspond to $\tilde R =0.5$ (blue), $\tilde R = 1$ (orange), and $\tilde R = 1.5$ (green). The horizontal dotted line marks equality with the ambient thermal bath temperature, $ T^\eff(E) =T$. The left panel shows the full range $0<v<1$, where the effective temperature is generally larger than the ambient temperature and grows rapidly as $v\to 1$ for each fixed $\tilde R$. At fixed $v$, $T^\eff(E)/T$ is larger for smaller $\tilde R$. The right panel zooms in on the small $v$ regime, $0<v \lesssim 0.32$, showing that each curve dips below the ambient value for sufficiently small $v$; the depth and width of this dip increase with $\tilde R$ for the curves shown.}
    \label{fig:small gap temp ratio}
\end{figure}

Figure \ref{fig:small gap temp ratio} shows the leading order term in \eqref{eq:beta T(E) small E} as a function of the orbital speed $v=\tilde \Omega \tilde R$ for several fixed values of $\tilde R$. For the values of $\tilde R$ shown, the effective temperature $T^\eff(E)$ is greater than the ambient bath temperature $T$ over most of the speed range. In particular, for $\tilde R \leq 1.5$, $T^\eff(E)>T$ for orbital speeds $v \gtrsim 0.3$, and grows rapidly as $v\to 1$. At lower speeds, $v\lesssim 0.3$, the effective temperature can fall slightly below the ambient temperature; however, for $\tilde R \lesssim 1.5$, this effect is small, with the minimum value remaining approximately above $T^\eff(E)\gtrsim 0.996 T$. The depth and width of the region in which $T^\eff(E)<T$ increases with $\tilde R$. However, in what follows we focus on values of $\tilde R$ for which the effective temperature is predominantly greater than the ambient bath temperature.

In Section \ref{sec: Otto CM small gap}, we shall also consider the derivative of $T^\eff(E)$ with respect to $E$ in the small $E$ limit. It was shown in \cite{Bunney:2023ipk,ParryThesis:2026} that the series expansion of $\widehat{\W}_T(E)$ to all orders may be written as $\widehat{\W}_T(E) = \alpha_0 - E/(4\pi) + \sum_{n=1}^\infty\alpha_n E^{2n}$ and furthermore that this series converges for $|E|<\gamma^3 \Omega^2 R$. An elementary calculation then shows that the temperature $T^\eff(E)$ has the expansion $T^\eff(E) = \sum_{n=0}^\infty\beta_n E^{2n}$, which is convergent for sufficiently small $|E|$. This then implies that $\frac{d}{dE}T^\eff(E) = O(E)$ as $E\to 0$.

\subsection{High ambient temperature limit}
Now consider the high ambient temperature limit $T\to \infty$ of \eqref{eq: CM thermal response} with $E$, $\Omega$ and $R$ fixed. The inertial vacuum contribution \eqref{eq:Fin(E)} is independent of~$T$, so it suffices to consider the $T\to \infty$ limit of $\widehat{\W}_T^\mathrm{corr}(E)$~\eqref{eq: Fcorr}. 
We parametrise this limit using $\tilde \Omega$ by writing $\tilde E=E\tilde \Omega/\Omega$ and $\tilde R = v/\tilde \Omega$. In terms of $\tilde \Omega$ \eqref{eq: Fcorr} becomes
\begin{equation} \label{eq: Fcorr omega tilde}
    \frac{1}{T}\widehat{\W}_T^\mathrm{corr}(E) =  \frac{1}{\pi \gamma} \intoinf \dd z \, \cos\!\left(\tfrac{2E}{\gamma \Omega}\tilde \Omega z \right)\!\left( \frac{\gamma^2}{z^2} - \frac{1}{2}\frac{\sinch\!\left( v\sinc\!\left(\tilde \Omega z \right)\!z\right)}{\cosh(z) - \cosh\!\left( v\sinc\!\left(\tilde \Omega z \right)\!z \right)} \right),
\end{equation}
where we have divided both sides by $T$.
This limit is then equivalent to the $\tilde \Omega \to 0$ limit  with $E$, $\Omega$ and $v$ fixed.

We show in Appendix \ref{app: DCT argument} that
\begin{equation} \label{eq: response high amb temp limit}
  \lim_{T\to \infty} \frac{1}{T}\widehat{\W}_T^\mathrm{corr}(E) = \frac{1}{4
    \pi\gamma v}\ln\!\left(\frac{1+v}{1-v} \right). 
\end{equation}
By combining \eqref{eq: response high amb temp limit} with \eqref{eq: CM thermal response} and \eqref{eq: effective temp}, we find that the ratio of the effective detailed balance temperature to the ambient thermal bath temperature, $T^\eff(E)/T$, in the large $T$ limit, is
\begin{equation} \label{eq:beta T(E) high T}
    \frac{T^\eff(E)}{T} = \frac{1}{2\gamma v}\ln\!\left(\frac{1+v}{1-v} \right)\Big(1 + o(1) \Big) . 
\end{equation}

\begin{figure}[t]
    \centering
    \includegraphics[width=0.6\linewidth]{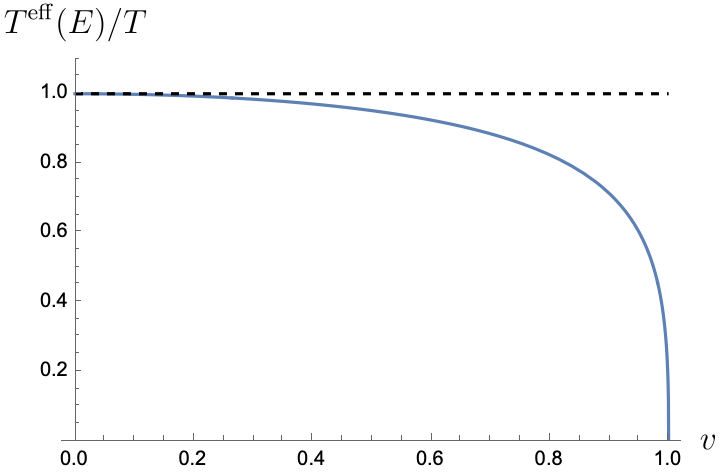}
    \caption{Leading order term \eqref{eq: high T leading order term} in the large $T$ expansion \eqref{eq:beta T(E) high T} of the temperature ratio, $T^\eff(E)/T$, as a function of the orbital speed $v$. The horizontal dotted line marks equality with the ambient temperature. }
    \label{fig:highambtemp}
\end{figure}

Two comments about \eqref{eq:beta T(E) high T} are in order. 
First, the leading order term in \eqref{eq:beta T(E) high T} is positive and  bounded above by unity, it is strictly decreasing in~$v$, and it has the limit $1$ as $v\to0_+$ and the limit $0$ as $v\to1_-$. These properties are seen by writing
\begin{equation} \label{eq: high T leading order term}
    \frac{1}{2\gamma v}\ln\!\left(\frac{1+v}{1-v} \right) = \frac{1}{\sinch\bigl(\tanh^{-1} (v)\bigr)}
\end{equation}
and using known properties of $\sinch$ and $\tanh^{-1}$. Thus, in the high ambient temperature limit, the effective temperature is lower than the ambient bath temperature for every $0<v<1$. Furthermore, the leading order temperature ratio \eqref{eq: high T leading order term} decreases as $v$ increases. Figure \ref{fig:highambtemp} shows the leading order high temperature ratio \eqref{eq: high T leading order term} as a function of the orbital speed~$v$, where the features we have mentioned can be readily seen. 

Second, the same leading order ratio appears for an inertial detector moving through a thermal bath at constant speed $v$~\cite{DM:OA:Efficiency}. This agreement has the following simple interpretation. The thermal correlations in the pulled-back Wightman function \eqref{eq: Wightman pullback} vary on the scale set by the inverse ambient temperature~$1/T$. In the high ambient temperature limit with fixed~$v$, this correlation scale becomes small compared with the timescale set by the circular motion, over which the circular trajectory is locally indistinguishable from an inertial trajectory with the same instantaneous speed relative to the bath.

\section{Quantum Otto cycle with circular motion}
\label{sec:otto-with-circular}

Having characterised the detector's response in circular motion in the preceding sections, we now investigate its implications for the performance of the quantum Otto thermal machine.

\subsection{Small gap limit} \label{sec: Otto CM small gap}

We first focus on the regime of small (but strictly positive) detector energy gaps, and analyse the performance of the Otto cycle operating either as a heat engine or a refrigerator. We assume that $E_c/E_h \to k$ with $0<k<1$ as $E_c\to 0$ and $E_h \to 0$.

\subsubsection{Heat engine}
\label{subsubsec:eff-at-max-power}


\begin{figure}
    \centering
    \includegraphics[width=0.34\linewidth]{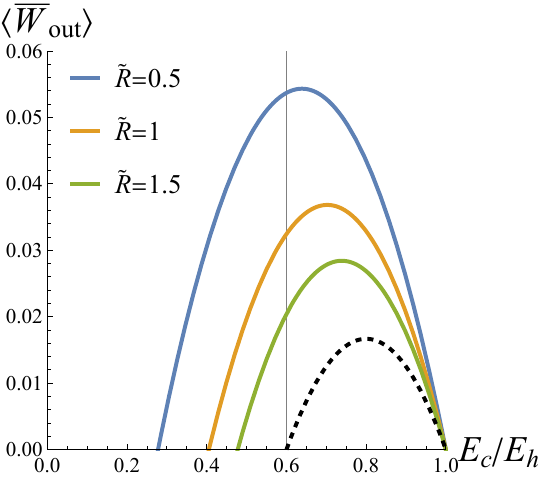}\hspace{0.2cm}\includegraphics[width=0.3\linewidth]{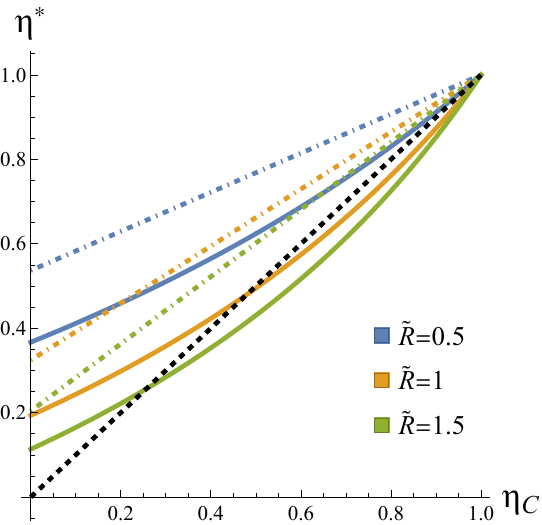}\hspace{0.2cm}\includegraphics[width=0.29\linewidth]{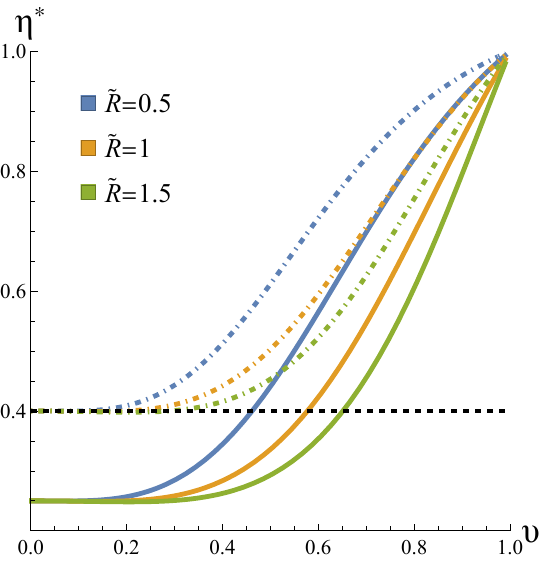}
    \caption{Small gap limit heat engine of Section~\ref{subsubsec:eff-at-max-power}, with the detector following a circular orbit during the hot isochore while remaining static during the cold isochore. \emph{Left}: Normalised work output, $\expval{\overline{W}_{\rm out}}=T_h\expval{W_{\rm out}}/E_h^2$, with $\expval{W_{\rm out}}$ given by \eqref{eq:work:small}, as a function of the gap ratio $E_c/E_h$, for a fixed bath temperature ratio $T_c/T_h=0.6$, and different values of $\tilde{R}$ at orbital speed $v=0.6$. The vertical line marks the boundary between refrigerator and heat engine operating regimes in the static case, determined by the positive work condition $E_c/E_h > T_c/T_h$. The black dotted line represents the work output in the static case ($v=0$). \emph{Centre}: Efficiency at maximum power (solid lines), given by \eqref{emp}, as a function of the Carnot efficiency, $\eta_C=1-T_c/T_h$, for $v=0.6$. The corresponding generalised Carnot bounds, $\eta_C^{\text{eff}}=1-T_c/T_h^{\text{eff}}(0)$, are shown as dot-dashed lines. \emph{Right}: Efficiency at maximum power, given by \eqref{emp}, as a function of the orbital speed $v$ for a fixed temperature ratio $T_c/T_h=0.6$. The dashed line indicates the standard Carnot efficiency, $\eta_C$, while the dot-dashed lines indicate the generalised Carnot bounds.} 
    \label{fig:smallgapperformance}
\end{figure}

First, consider the heat engine case. We take the detector to be in uniform circular motion during the hot isochore where it perceives an effective temperature $T^\eff_h(E_h)$, and we take the detector to be static during the cold isochore where it thermalises to the ambient temperature $T_c$. Recall that the leading order small-gap expansion of $T^\eff_h(E_h)/T_h$ is given by \eqref{eq:beta T(E) small E}. We shall be interested in the parameter regimes, identified in Section \ref{sec: CM small gap}, in which $T^\eff_h(E_h)>T_h$.

We characterise the performance of the heat engine by the EMP $\eta^*$ described in Section \ref{sec: EMP}. Recall from Section \ref{sec:quantum-otto} that the total work output over one cycle is  $\expval{W_\mathrm{out}} = -\Delta\!\expval{W}$ where $\Delta\!\expval{W}$ is given by \eqref{eq: net work} and the power is given by \eqref{eq: power}. In the small gap limit, $\expval{W_\mathrm{out}}$ is approximately given by
\begin{equation}\label{eq:work:small}
    \expval{W_\mathrm{out}} = \frac{E_h-E_c}{4}\!\left(\frac{E_c}{T_c} - \frac{E_h}{T^\eff_h(0)}\right).
\end{equation}

Now we compute the EMP in the small gap limit. The condition $\partial P / \partial E_h = 0$, where $P$ is given by \eqref{eq: power}, gives 
\begin{align}
    0 &= -\frac{1}{2}\!\left( \tanh\!\left(\frac{E_h}{2T^\eff_h(E_h)} \right) -\tanh\!\left(\frac{E_c}{2T_c} \right)  \right) \notag \\
    &\hspace{0.4cm}+ \frac{E_c-E_h}{4[T^\eff_h(E_h)]^2} \left( T^\eff_h(E_h) - E_h \frac{d}{dE_h}T^\eff_h(E_h) \right)\sech^2\!\left(\frac{E_h}{2T^\eff_h(E_h)} \right).
\end{align}
Subsequently, taking both $E_c$ and $E_h$ to be small such that $E_c/E_h \to k$ with $0<k<1$, we find
\begin{equation}\label{eq: EMP E_c/E_h}
     \frac{E_c}{E_h} = \frac{2}{1 + T^\eff_h(0)/T_c} + O(E_c^2+E_h^2),
\end{equation}
where $T^\eff_h(0) = \lim_{E_h\to 0}T^\eff_h(E_h)$, and where we have used that $\frac{d}{dE_h}T^\eff_h(E_h)= O(E_h)$ as $E_h\to 0$. Substitution of \eqref{eq: EMP E_c/E_h} into the engine efficiency \eqref{eq: efficiency 2} then gives the EMP
\begin{align}\label{emp}
    \eta^*=\frac{\eta_C^{\text{eff}}}{2-\eta_C^{\text{eff}}} + O(E_c^2+E_h^2),
\end{align}
where the generalised Carnot efficiency reads $\eta_C^{\text{eff}}=1-T_c/T_h^{\text{eff}}(0)$. In the absence of motion, \eqref{emp} reduces to the efficiency at maximum power
of a two-level quantum Otto engine \cite{singh:OA}, with the resulting  expression, $\eta^*=\eta_C/2+\mathcal{O}(\eta_C^2)$ 
agreeing to leading order with the expansion of the Curzon–Ahlborn efficiency  \cite{Curzon:Ahlborn}.

Figure \ref{fig:smallgapperformance} (left panel) shows the normalised total work output, $\expval{\overline{W}_{\rm out}}=T_h\expval{W_{\rm out}}/E_h^2$, as a function of the energy gap ratio $E_c/E_h$ for a fixed value of the temperature ratio $T_c/T_h=0.6$ and orbital speed $v=0.6$, and for several values of~$\tilde{R}$. For comparison, we also include the standard Otto cycle, in which the working medium remains static during both the isochoric strokes. We observe that the generalised positive work condition, $E_c/E_h>T_c/T_h^{\text{eff}}(0)$, extends the heat-engine operating regime to parameters for which the corresponding static cycle would operate as a refrigerator. In particular, for $T_c/T_h=0.6$, $v=0.6$, and $\tilde R=1$, the cycle operates as a heat engine for $E_c/E_h\gtrsim 0.4$, whereas the static cycle requires $E_c/E_h>0.6$.
The centre and right panels of Figure \ref{fig:smallgapperformance} show the EMP, given by \eqref{emp}, as a function of the standard Carnot efficiency $\eta_C$ and the orbital speed $v$, respectively. We observe that, for the different values of $\tilde R$ considered, the EMP can exceed the standard Carnot efficiency, defined in terms of the rest-frame bath temperatures. In particular, for $\tilde R=0.5$, the EMP exceeds the standard Carnot bound already at orbital speeds above $v\approx0.45$. Nevertheless, the EMP always remains below the generalised Carnot efficiency, $\eta_C^{\text{eff}}=1-T_c/T_h^{\text{eff}}(0)$, which account for the effective hot bath temperature experienced by the moving working medium.

\subsubsection{Refrigerator}
\label{subsubsec:max-figure-merit}

Second, consider the refrigerator case. Here we take the detector to be in uniform circular motion during the cold isochore where it perceives an effective temperature $T^\eff_c(E_c)$, and we take the detector to be static during the hot isochore where it thermalises to the ambient temperature $T_h$. Again, the leading order small gap expansion of $T^\eff_c(E_c)/T_c$ is given by \eqref{eq:beta T(E) small E}, and we shall be interested in the parameter regimes, identified in Section \ref{sec: CM small gap}, in which $T^\eff_c(E_c)>T_c$.

We characterise the performance of the refrigerator by the COP at  maximum figure of merit $\varepsilon^*$ described in Section \ref{sec: COP maximum figure}. Recall that the heat extracted from the cold reservoir $\expval{Q_c}>0$ is given by \eqref{eq: heat DA} and the figure of merit $\chi_\mathrm{merit}$ is given in \eqref{eq: figure of merit}.
In the small gap limit, $\expval{Q_c}$ is approximately given by
\begin{equation}\label{eq:qc}
    \expval{Q_c} = \frac{E_c}{4}\!\left(\frac{E_h}{T_h} - \frac{E_c}{T^\eff_c(0)} \right).
\end{equation}

Now we compute the COP at maximum figure of merit in the small gap limit. The condition $\partial \chi_\mathrm{merit} / \partial E_c=0$ then gives
\begin{align}
    0 &= \frac{E_c(2E_h-E_c)}{2}\!\left(\tanh\!\left(\frac{E_h}{2T_h} \right) -\tanh\!\left(\frac{E_c}{2T^\eff_c(E_c)} \right)\right) \notag \\
    &\hspace{0.4cm}- \frac{E_c^2(E_h-E_c)}{4[T^\eff_c(E_c)]^2}\left( T^\eff_c(E_c) - E_c \frac{d}{dE_c}T^\eff_c(E_c) \right)\sech^2\!\left(\frac{E_c}{2T^\eff_c(E_c)} \right).
\end{align}
Subsequently, taking both $E_c$ and $E_h$ to be small such that $E_c/E_h\to k$ with $0<k<1$, we find
\begin{equation} \label{eq: E_c/E_h fridge expansion}
    \frac{E_c}{E_h} = \frac{1}{4}\left[3 + \frac{T^\eff_c(0)}{T_h} - \sqrt{\left(\frac{T^\eff_c(0)}{T_h}-1 \right)\left(\frac{T^\eff_c(0)}{T_h}-9 \right)} \right] + O(E_c^2+E_h^2),
\end{equation}
where $T^\eff_c(0) = \lim_{E_c\to 0}T^\eff_c(E_c)$, and where we have used $\frac{d}{dE_c}T^\eff_c(E_c)= O(E_c)$ as $E_c\to 0$. Substitution of \eqref{eq: E_c/E_h fridge expansion} into the COP \eqref{eq: COP 2} then gives the COP at maximum figure of merit
\begin{align}\label{COP:max:chi}
\varepsilon^*=\frac{1}{2}\left(\sqrt{8\varepsilon_C^{\text{eff}}+9}-3\right) + O(E_c^2 + E_h^2),
\end{align}
where the generalised Carnot COP reads $\varepsilon_C^\eff = (T_h/T^\eff_c(0)-1)^{-1}$. In the absence of motion, this expression recovers the known result for a two-level quantum Otto refrigerator~\cite{singh:OA}.

\begin{figure}
    \centering    \includegraphics[width=0.45\linewidth]{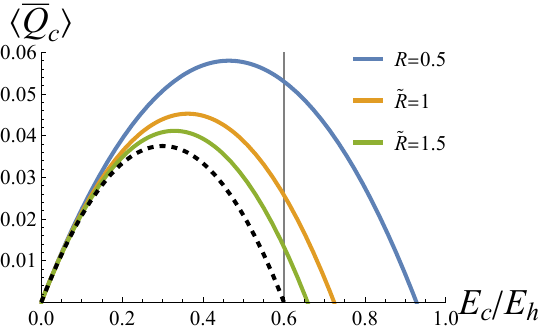}\hspace{1cm}\includegraphics[width=0.4\linewidth]{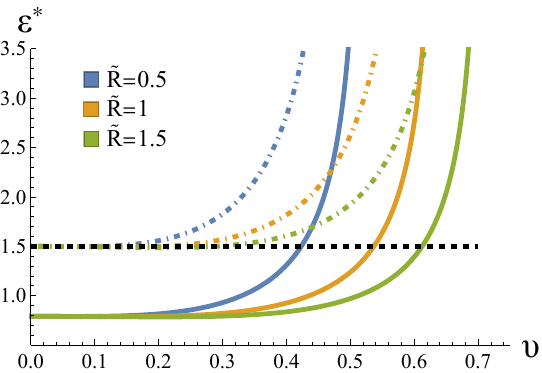}
    \caption{Small gap limit refrigerator of Section~\ref{subsubsec:max-figure-merit}, with the detector following a circular orbit during the cold isochore while remaining static during the hot isochore. \emph{Left}: Normalised heat extracted from the cold bath, $\expval{\overline{Q}_c}=T_h\expval{Q_c}/E_h^2$, with $\expval{Q_c}$ given by \eqref{eq:qc}, as a function of the gap ratio $E_c/E_h$ for a fixed bath temperature ratio $T_c/T_h=0.6$ and different values of $\tilde{R}$ at orbital speed $v=0.5$. The vertical line marks the boundary between refrigerator and heat engine operating regimes in the static case, determined by the refrigeration condition $E_c/E_h < T_c/T_h$. The black dotted line corresponds to the static case ($v=0$). \emph{Right}: COP at maximum figure of merit, given by \eqref{COP:max:chi}, as a function of the orbital speed $v$ for a fixed temperature ratio $T_c/T_h=0.6$. The dashed line indicates the standard Carnot COP, $\varepsilon_C=(T_h/T_c-1)^{-1}$, while the dot-dashed lines indicate the generalised Carnot bound $\varepsilon_C^\eff = (T_h/T^\eff_c(0)-1)^{-1}$.} 
    \label{fig:smallgapperformance:refrig}
\end{figure}

Figure \ref{fig:smallgapperformance:refrig} (left panel) shows the normalised heat extracted from the cold bath, $\expval{\overline{Q}}_c=T_h\expval{Q_c}/E_h^2$, as a function of the energy gap ratio $E_c/E_h=$, for a fixed value of the temperature ratio $T_c/T_h=0,6$ and orbital speed $v=0.5$, and for several values $\tilde{R}$. As in the heat engine case, motion-induced changes in the effective temperature modify the operating regime, allowing refrigeration under conditions for which the corresponding static cycle would not provide cooling. In particular, for $T_c/T_h=0.6$, $v=0.5$, and $\tilde R=1$, the generalised refrigeration condition implies that the cycle operates as a refrigerator for $E_c/E_h\lesssim0.72$, whereas the static cycle requires $E_c/E_h<0.6$.  The right panel shows the COP at maximum figure of merit, given by \eqref{COP:max:chi}, as a function of the orbital speed $v$. We observe that the COP can exceed the standard Carnot bound, defined in terms of the rest-frame bath temperatures, while remaining below the generalised Carnot COP, $\varepsilon_C^\eff = (T_h/T^\eff_c(0)-1)^{-1}$, which accounts for the effective temperature experienced by the moving working medium during the cold isochore.

These findings bear a close analogy to quantum Otto engines and refrigerators coupled to engineered squeezed thermal reservoirs \cite{Huang:2012,OA:2014,Long:2015}. In such systems, squeezing modifies the stationary state of the working medium and can, under suitable conditions, be described by an effective temperature exceeding that of the thermal reservoir. As a result, the efficiency and COP may surpass the standard Carnot expressions defined in terms of the reservoir temperatures. Although squeezing and relative motion arise from distinct physical mechanisms, both can modify the effective thermal response of the working medium and thereby influence its operating regime and thermodynamic performance. In particular, squeezing involves the preparation of a nonequilibrium reservoir, whereas relative motion modifies the detector's response to an otherwise equilibrium thermal field.

\subsection{High ambient temperature limit}
\label{subsec:high-ambient-temperature-engine}

In the high ambient temperature limit, the effective temperature \eqref{eq:beta T(E) high T} experienced by the detector becomes lower than the bath temperature, and the circular motion response approached that of an inertial detector moving at constant velocity through a thermal bath. In this regime, quantum Otto heat engines and refrigerators can also be realised by allowing the detector to move only during the cold isochoric stroke of the heat engine cycle and the hot isochoric stroke of the refrigeration cycle.

\begin{figure}
    \centering    \includegraphics[width=0.4\linewidth]{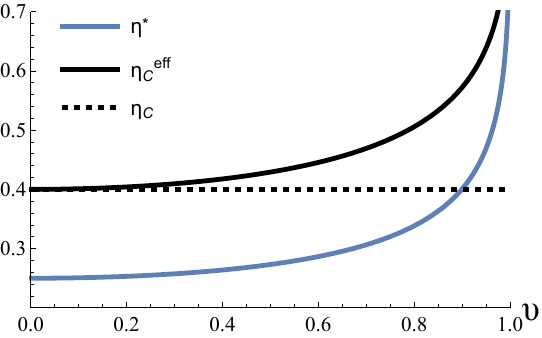}\hspace{1cm}\includegraphics[width=0.4\linewidth]{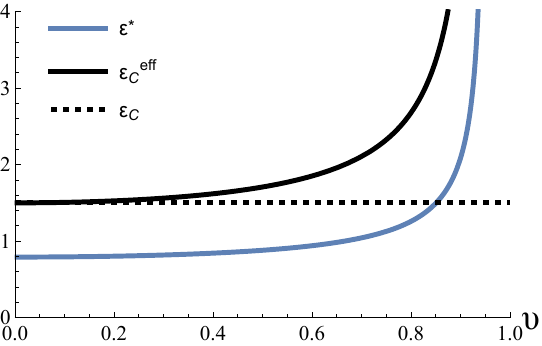}
    \caption{High ambient temperature scenario of Section~\ref{subsec:high-ambient-temperature-engine}. \emph{Left}: 
    Efficiency at maximum power of the heat engine, given by \eqref{emp} with $\eta_C^{\text{eff}}=1-T_c^{\text{eff}}(0)/T_h$. \emph{Right}: coefficient of performance at maximum figure of merit given by \eqref{COP:max:chi} with $\varepsilon_C^\eff = (T_h^\eff(0)/T_c-1)^{-1}$ of the refrigerator. Both quantities are shown as functions of the constant velocity $\upsilon$ of the inertially moving detector for a fixed bath temperature ratio $T_c/T_h=0.6$.}
    \label{fig:inertial}
\end{figure}

As demonstrated in \cite{DM:OA:Efficiency,DM:OA:TUR}, the efficiency at maximum power and the COP at maximum figure of merit can exceed the standard Carnot expressions defined in terms of the rest-frame bath temperatures, while remaining bounded by their generalised counterparts based on the effective temperatures experienced by the moving detector. For completeness, Figure \ref{fig:inertial} presents these optimised performance measures as functions of the detector velocity. Comparison with the circular motion results reveals that, for the same bath temperature ratio, substantially higher velocities are required in the inertial case to exceed the conventional Carnot values.

\section{Closing remarks}
\label{sec:closing-remarks}

In this work, we have investigated the response of a qubit Unruh–DeWitt detector undergoing uniform circular motion in (3+1)-dimensional Minkowski spacetime, with the quantum field prepared in a thermal state. We have characterised the detector's effective temperature in the small-energy-gap and high-ambient-temperature regimes. These results were subsequently applied to a quantum Otto heat engine and a refrigerator to examine how motion-induced effective temperatures influence their operating conditions and performance.

We have shown that circular motion can extend the operating regimes of both machines, allowing their performance to exceed the conventional Carnot values defined in terms of the rest-frame bath temperatures, while remaining bounded by their generalised counterparts based on the effective temperatures. In the high-ambient-temperature regime, the circular-motion response approaches that of an inertially moving detector. While inertial motion through a thermal bath modifies the observed spectrum through relativistic Doppler shifts, circular motion introduces additional acceleration-dependent effects that influence the detector's effective temperature and, consequently, the performance of the quantum thermal machine.

Our findings highlight how relative motion modifies the operating regimes and performance bounds of quantum thermal machines. In this sense, relativistic motion can act as an effective thermodynamic resource (see, also, \cite{DM:OA:Efficiency,DM:OA:TUR,pandit2025,shastri2025}), as a detector moving through an equilibrium thermal field can exhibit a non-thermal response, characterised by an energy gap dependent effective temperature. Unlike engineered nonequilibrium reservoirs, where the thermal properties are modified through reservoir preparation, these effects arise from the detector's motion relative to the field.

\section*{Acknowledgments}

DM was supported by the UK Research and
Innovation Engineering and Physical Sciences Research Council (Grant No. EP/Z002796/1).
BAJ-A is supported by the EPSRC Open
Fellowship EP/Y014510/1. 
The work of JL was supported by United Kingdom Research and Innovation Science and Technology Facilities Council [grant numbers ST/S002227/1, ST/T006900/1, ST/Y004523/1 and ST/C004235/1].
The authors have benefited from the activities of COST Action CA23115:
Relativistic Quantum Information, funded  by COST (European Cooperation in Science and Technology). 
For the purpose of open access, the authors have applied a CC BY public copyright licence to any Author Accepted Manuscript version arising. 


\appendix

\section{High temperature limit of 
$\frac{1}{T} \widehat{\W}_T^{\mathrm{corr}}(E)$ \eqref{eq: Fcorr omega tilde}}
\label{app: DCT argument}

In this appendix, we compute the high temperature limit of $\frac{1}{T} \widehat{\W}_T^{\mathrm{corr}}(E)$ \eqref{eq: Fcorr omega tilde}, 
\begin{equation} \label{eq: Fcorr omega tilde app}
    \frac{1}{T} \widehat{\W}_T^{\mathrm{corr}}(E) =  \frac{1}{\pi \gamma} \intoinf \dd z \, \cos\!\left(k\tilde \Omega z \right)\!\left( \frac{\gamma^2}{z^2} - \frac{1}{2}\frac{\sinch\!\left( v\sinc\!\left(\tilde \Omega z \right)\!z\right)}{\cosh(z) - \cosh\!\left( v\sinc\!\left(\tilde \Omega z \right)\!z \right)} \right),
\end{equation}
with $v$ and $k$ fixed, where $k=2E/(\gamma \Omega)$, and the high temperature limit is the limit $\tilde\Omega \to 0$. In particular, we shall construct a dominated convergence argument that allows us to interchange the pointwise $\tilde \Omega \to 0$ limit and the integral in \eqref{eq: Fcorr omega tilde app}.

First, let
\begin{equation}
    I(z,\tilde\Omega) := \frac{\gamma^2}{z^2} - \frac{\sinch\!\left( v \sinc(\tilde \Omega z)z \right)}{2\left[ \cosh(z) - \cosh\!\left( v \sinc(\tilde \Omega z)z \right) \right]}, 
\label{eq:F-def}
\end{equation}
where $z>0$ and $\tilde \Omega >0$. Note that $I(z,\tilde \Omega)$ is the expression in the large parentheses in the integrand in~\eqref{eq: Fcorr omega tilde app}, with the dependence on $\tilde \Omega$ indicated explicitly. The pointwise $\tilde \Omega \to 0$ limit of $I(z,\tilde \Omega)$ is
\begin{equation}
I(z,\tilde\Omega) 
\xrightarrow[\tilde\Omega\to0]{}
    \frac{\gamma^2}{z^2} - \frac{\sinch(vz)}{2[\cosh(z)-\cosh(vz)]} =: I(z,0), 
\end{equation}
and we have $I(z,0)>0$, using $0<v<1$ and the estimate 
\begin{equation}
\cosh(z)-\cosh(vz) 
= z\int_v^1 \dd u \, \sinh(z u) 
= z^2\int_v^1 \dd u \, u \sinch(z u) 
> z^2 \sinch(z v )  \int_v^1 \dd u \, u 
= \frac{z^2}{2\gamma^2} \sinch(z v ) . 
\end{equation}

When $z\ge1$, using $|\sinc(x)|\le1$ and the triangle inequality in \eqref{eq:F-def} we obtain the bound 
\begin{equation}
    |I(z,\tilde\Omega)| \le \frac{\gamma^2}{z^2} + \frac{\sinch\!\left( v z \right)}{2\left[ \cosh(z) - \cosh\!\left( v z \right) \right]} . 
\end{equation}
When $0<z\le1$ and $\tilde\Omega\le\pi$, an elementary analysis shows that $\partial_{\tilde\Omega}I(z,\tilde\Omega) \ge 0$, which gives the bound 
$0 \le I(z,\tilde\Omega) \le I(z,\pi)$. 

Combining, it follows that when $\tilde\Omega\le\pi$, we have 
\begin{align}
|I(z,\tilde\Omega)| \le 
\begin{cases}
{\displaystyle{\frac{\gamma^2}{z^2} + \frac{\sinch\!\left( v z \right)}{2\left[ \cosh(z) - \cosh\!\left( v z \right) \right]}}}
& \text{for}\ z\ge1, 
\\
I(z,\pi) 
& \text{for}\ 0 < z \le1, 
\end{cases}
\end{align}
where the right-hand side is independent of $\tilde\Omega$ and integrable over $(0,\infty)$. 
As the integrand in \eqref{eq: Fcorr omega tilde app} is bounded in absolute value by $|I(z,\tilde\Omega)|$, it follows by dominated convergence that the $\tilde \Omega \to 0$ limit of \eqref{eq: Fcorr omega tilde app} can be taken under the integral, with the outcome
\begin{equation} \label{eq:high_temp_I-limit}
     \lim_{\tilde \Omega \to 0} \frac{1}{T} \widehat{\W}_T^\mathrm{corr}(E)
     = \frac{1}{2\pi}J(v), 
\end{equation}
where 
\begin{equation} \label{eq:J(v)}
     J(v) = \frac{2}{\gamma} \int_0^\infty \dd z \, \left( \ \frac{\gamma^2}{z^2} - \frac{\sinch(vz)}{2(\cosh(z)-\cosh(vz))} \right) . 
\end{equation}

We may evaluate the integral in \eqref{eq:J(v)} in terms of known functions. An elementary application of trigonometric identities shows that $J(v)$ can be recast as 
\begin{equation}\label{eq: Fcorr beta->0 2}
     J(v) =  \frac{1}{2\gamma v } \intoinf \dd z \, \frac{f\big(\tfrac{1}{2}(1+v)z \big) - f\big(\tfrac{1}{2}(1-v)z \big)}{z},
\end{equation}
where $f(u) = \coth(u) -1/u$ is $C^1(\mathbb{R})$ with limits $\lim_{u\to 0}f(u)=0$ and $\lim_{u\to \infty}f(u)=1$. The integral in \eqref{eq: Fcorr beta->0 2} evaluates to $\ln\!\left(\frac{1+v}{1-v} \right)$ by Frullani's theorem \cite{Tricomi01031951}, and hence
\begin{equation} \label{eq:J(v)2}
    J(v) = \frac{1}{2\gamma v}\ln\!\left(\frac{1+v}{1-v} \right).
\end{equation}

Combining, we find that 
\begin{equation}
    \lim_{T\to \infty}\frac{1}{T} \widehat{\W}_T^\mathrm{corr}(E) = \frac{1}{4
    \pi\gamma v}\ln\!\left(\frac{1+v}{1-v} \right).
\end{equation}

\printbibliography
\end{document}